
%



\font\rmu=cmr10 scaled\magstephalf
\font\bfu=cmbx10 scaled\magstephalf

\font\it=cmti10 scaled \magstephalf
\font\bf=cmbx10 scaled\magstephalf
\rmu



\font\rmus=cmr8
\font\rmuss=cmr6

\font\mait=cmmi10 scaled\magstephalf
\font\maits=cmmi7 scaled\magstephalf
\font\maitss=cmmi7

\font\msyb=cmsy10 scaled\magstephalf
\font\msybs=cmsy8 scaled\magstephalf
\font\msybss=cmsy7

\font\bfus=cmbx7 scaled\magstephalf
\font\bfuss=cmbx7

\font\cmeq=cmex10 scaled\magstephalf

\textfont0=\rmu
\scriptfont0=\rmus
\scriptscriptfont0=\rmuss

\textfont1=\mait
\scriptfont1=\maits
\scriptscriptfont1=\maitss

\textfont2=\msyb
\scriptfont2=\msybs
\scriptscriptfont2=\msybss

\textfont3=\cmeq
\scriptfont3=\cmeq
\scriptscriptfont3=\cmeq

\newfam\bmufam  \textfont\bmufam=\bfu
      \scriptfont\bmufam=\bfus \scriptscriptfont\bmufam=\bfuss


\hsize=15.5cm
\vsize=22cm
\baselineskip=16pt   
\parskip=16pt plus  2pt minus 2pt




\def\ni{\noindent}

\def\a{\alpha}

\def\d{\delta}

\def\f{\phi}
\def\g{\gamma}
\def\l{\lambda}

\def\ut#1 {\rlap{\lower1ex\hbox{$\sim$}}#1{}}

\def\R{{\rm I\!R}}

\def\Q{{\mathchoice
{\setbox0=\hbox{$\displaystyle\rm Q$}\hbox{\raise 0.15\ht0\hbox to0pt
{\kern0.4\wd0\vrule height0.8\ht0\hss}\box0}}
{\setbox0=\hbox{$\textstyle\rm Q$}\hbox{\raise 0.15\ht0\hbox to0pt
{\kern0.4\wd0\vrule height0.8\ht0\hss}\box0}}
{\setbox0=\hbox{$\scriptstyle\rm Q$}\hbox{\raise 0.15\ht0\hbox to0pt
{\kern0.4\wd0\vrule height0.7\ht0\hss}\box0}}
{\setbox0=\hbox{$\scriptscriptstyle\rm Q$}\hbox{\raise 0.15\ht0\hbox to0pt
{\kern0.4\wd0\vrule height0.7\ht0\hss}\box0}}}}
\def\C{{\mathchoice
{\setbox0=\hbox{$\displaystyle\rm C$}\hbox{\hbox to0pt
{\kern0.4\wd0\vrule height0.9\ht0\hss}\box0}}
{\setbox0=\hbox{$\textstyle\rm C$}\hbox{\hbox to0pt
{\kern0.4\wd0\vrule height0.9\ht0\hss}\box0}}
{\setbox0=\hbox{$\scriptstyle\rm C$}\hbox{\hbox to0pt
{\kern0.4\wd0\vrule height0.9\ht0\hss}\box0}}
{\setbox0=\hbox{$\scriptscriptstyle\rm C$}\hbox{\hbox to0pt
{\kern0.4\wd0\vrule height0.9\ht0\hss}\box0}}}}

\font\fivesans=cmss10 at 4.61pt
\font\sevensans=cmss10 at 6.81pt
\font\tensans=cmss10
\newfam\sansfam
\textfont\sansfam=\tensans\scriptfont\sansfam=\sevensans\scriptscriptfont
\sansfam=\fivesans
\def\sans{\fam\sansfam\tensans}
\def\Z{{\mathchoice
{\hbox{$\sans\textstyle Z\kern-0.4em Z$}}
{\hbox{$\sans\textstyle Z\kern-0.4em Z$}}
{\hbox{$\sans\scriptstyle Z\kern-0.3em Z$}}
{\hbox{$\sans\scriptscriptstyle Z\kern-0.2em Z$}}}}
\def\semi{\bigcirc\kern-1em{s}\;}



\newcount\foot
\foot=1
\def\note#1{\footnote{${}^{\number\foot}$}{\ftn #1}\advance\foot by 1}

\def\frac#1#2{{#1\over #2}}
\def\text#1{\quad{\hbox{#1}}\quad}


\font\ch=cmbx12 scaled\magstephalf
\font\ftn=cmr8 scaled\magstephalf

\font\it=cmti10 scaled\magstephalf
\font\bf=cmbx10 scaled\magstephalf

\vbox{ {}\vskip2cm\vfill}

\vbox{\baselineskip=15 pt
 \vskip-0.9in
 \hfill \vbox{
  }}
\vskip.3cm

\centerline{\bf GENERALIZED COORDINATES ON THE PHASE}
\centerline{\bf SPACE OF YANG-MILLS THEORY}

\vskip 2cm

\centerline{R. Loll and J.M. Mour\~ao\note{On leave of absence from Dept.
F\'{\i}sica, Inst. Sup. T\'{e}cnico, 1096 Lisboa, PORTUGAL}}

\centerline{\it Center for Gravitational Physics and Geometry}
\centerline{\it Physics Department, Pennsylvania State University}
\centerline{\it University Park, PA 16802-6300, USA}

\bigskip

\centerline{J.N. Tavares}

\centerline{\it Dep. Matem\'atica Pura, Faculdade de Ci\^{e}ncias}
\centerline{\it Univ. Porto, 4000 Porto, PORTUGAL}
\vskip2cm

\centerline{\bf Abstract}

We study the suitability of complex Wilson loop variables
as (generalized) coordinates on the physical phase space
of $SU(2)$-Yang-Mills theory. To this end, we construct a natural one-to-one
map from the physical phase space of the Yang-Mills theory with compact gauge
group $G$ to a subspace of the physical configuration space of the complex
$G^\C$-Yang-Mills theory.
Together with a recent result by Ashtekar and Lewandowski this implies that
the complex Wilson loop variables form a complete set of generalized
coordinates
on the physical phase space of $SU(2)$-Yang-Mills theory. They also form a
generalized canonical loop algebra. Implications for both general relativity
and gauge theory
 are discussed.

\vskip 1cm

PACS numbers: 0240, 0460, 1190

\vfill


\eject

\line{\ch 1. Introduction\hfil}

The $SL(2,\C)$-Ashtekar connection introduced in [1] has led to important
progress in the Hamiltonian formulation of general relativity [2-4].
In particular a wide class of solutions to the quantum gravitational
equations has been found with the help of the Wilson
loop variables constructed from this connection [3,4].

The situation we encounter in the gravitational application is similar to
the gauge theory case to be discussed below. Namely, one has the
$SL(2,\C)$-Ashtekar connection  $A_{\rm grav}^\C$ on a spatial manifold
$\Sigma, \ dim(\Sigma ) = 3$,

$$
A_{\rm grav}^\C{}^j_a(x)=\Gamma_a^j (x)-i\, K_a^j(x),\eqno(1.1)
$$

\ni together with its complex conjugate, to be thought of as complex
coordinates on the real phase space of general relativity. In (1.1),
$\Gamma_a^i$ is the spin connection determined by the triad $E^a_i=
\frac{1}{\sqrt{h}}\tilde E^a_i$, and $K_a^i$ is related to the extrinsic
curvature $K_{ab}$ on $\Sigma$ by $K_a^i=\frac{1}{\sqrt{h}}K_{ab} E^{bi}$
[1,2]. Both $\Gamma$ and $K$ are coordinates on phase space,
but unlike in the Yang-Mills case we will consider, they are not
canonically conjugate to each other.

The Wilson loop variables constructed from the connections (1.1) too are
functions on the phase space of general relativity, invariant with respect
to $SL(2,\C)$-gauge transformations. Since physically one requires only
invariance under gauge transformations corresponding
to a $SU(2)$-subgroup of $SL(2,\C)$ [2], it may happen that the complex,
$SL(2,\C)$-invariant Wilson
loop variables do not separate points in the ``physical" phase space. (Here we
mean the phase space obtained after enforcing the reality conditions and
the Gauss law constraint but prior to imposing the diffeomorphism
and hamiltonian constraints.) Such pathologies may occur in the course of
going to the quotient space with respect to the $SL(2,\C)$-transformations,
although,
as was shown in [5], the Wilson loops separate all (separable)
points in the space of $SL(2,\C)$-connections modulo
$SL(2,\C)$-gauge transformations.

One finds an analogous situation, but with a much simpler symplectic
structure, in $SU(2)$ (or in general $G$)-Yang-Mills theories in a
$d+1$-dimensional space-time $M= \R \times \Sigma$. In the
present paper we prove that, for any compact gauge group $G$,
the physical phase space ${\cal P}_{\rm phys}$ of the $G$-Yang-Mills theory
can be naturally identified with a subset of ${\cal Q}^\C_{\rm phys}$,
the physical {\it configuration} space of the $G^\C$-Yang-Mills theory
(with $G^\C$ denoting the complexification of the Lie group $G$).
This is the case although the group ${\cal G}^\C$ of complex $G^\C$-gauge
transformations acting on the big phase space $\cal P$ is ``twice as large" as
the group $\cal G$ of real $G$-gauge transformations. Locally,
the identification between the space ${\cal P}_{\rm phys}$ and (a subspace of)
${\cal Q}^\C_{\rm phys}$ is made possible by the fact that the
imaginary $G^\C$-gauge transformations are transverse in the phase space
$\cal P$ to
the
constraint surface $\cal C$, defined by the vanishing of the non-linear
Gauss law constraints,

$$
{\cal C} = \{ (A_a, \tilde E^a) \in {\cal P} \ : \ D[A]_a \tilde E^a = 0\},
\eqno(1.2)
$$

\ni where $A_a$ is a $G$-gauge field, $\tilde E^a$ is a one-density
electric field and $D[A]_a$
 is the covariant derivative $D[A]_a \equiv D_a = \partial_a + [A_a,
\cdot ]$. Our proof however will be of a global nature.

For any closed piecewise smooth spatial curve $\alpha$, consider the complex
Wilson loop variable

$$
T_\a (A_a + i \l E_a) := \frac{1}{N}\,Tr\,{\rm P}\, \exp \left(\int_\a (A_a + i
\l E_a ) dx^a \right),
\eqno(1.3)
$$

\ni where $A_a\in {\cal A}^G$ is a $G$-connection and $E_a(x)$ is defined by
$E_a(x) = h_{ab}(x) {1 \over \sqrt{h}} \tilde E^b(x) $. The normalization
factor on the right-hand side is the dimension $N$ of the linear
representation of $G$, $h_{ab}$ is a fixed Riemannian metric on the spatial
manifold $\Sigma$,  and
$\l >0$ a positive real constant with dimension of length.
By virtue of their local
$G^\C$-gauge invariance, variables of the form (1.3) project down to the
physical  configuration space ${\cal Q}^\C_{\rm phys}={\cal A}^{G^\C}/{\cal
G}^{G^\C}$ of $G^\C$-Yang-Mills theory.

Moreover, it was shown
in [5] that for $G^\C = SL(2, \C)$ they constitute a complete set of
generalized coordinates on ${\cal
Q}^\C_{\rm phys}$ (i.e. they separate all points that are separable with
the help of  continuous functions).
Our result, for the particular case of $G = SU(2)$, and the
result of [5] imply
that the complex variables $T_\a$ can be used as global generalized
coordinates
on the physical phase space ${\cal P}_{\rm phys}$ of the $SU(2)$-Yang-Mills
theory (see comment after (2.9)). By ``generalized" we mean to indicate that,
due to the non-linearity
of the spaces ${\cal A}/ {\cal G}$, the Wilson loop variables are always
overcomplete,
subject to a set of identities, the so-called Mandelstam constraints [6,7].

\vskip 1.5cm

\line{\ch 2. Physical phase space of real Yang-Mills theory versus
physical\hfil}
\line{\ch \hskip0.7cm configuration space of complex Yang-Mills theory\hfil}

Consider the Yang-Mills theory with compact gauge group $G$ in a
$d+1$-dimensional space-time $\Sigma\times\R$. The canonical pairs

$$
(A_a(x), \tilde E^a(x)),
$$

\ni are a natural set of coordinates on phase space, with Poisson brackets
given by

$$
\lbrace A_a(x), \tilde E^b(y) \rbrace = \delta_a^b\delta^d(x,y). \eqno(2.1)
$$

\ni Note that the parametrization of the phase space
with $(A, \tilde E)$ is valid globally
whenever $dim (\Sigma) =3$
and $G=SU(2)$, because then the underlying principal fibre bundle is trivial.
The action of the group of gauge transformations $\cal G$,

$$
\eqalign{
A_a(x) &\mapsto  g^{-1}(x) A_a(x) g(x) + g^{-1}(x) \partial_a g(x) \cr
\tilde E^a(x) &\mapsto  g^{-1}(x) \tilde E^a(x) g(x),}\eqno(2.2)
$$

\ni can be naturally extended to an action of the complexified group
${\cal G}^\C$ if we identify the {\it big phase space} ${\cal P}$
of the $G$-Yang-Mills theory with the {\it big configuration space}
${\cal Q}^\C$ of the $G^\C$-Yang-Mills theory through the
($\lambda$-dependent) map

$$
\eqalignno{
\phi_\l \ &: \ {\cal P}  \ \rightarrow \ {\cal Q}^\C  \cr
\phi_\l \ & : \ \left(A_a(x), \tilde E^a(x)\right)  \
\mapsto \  A^\C_a(x) =
A_a(x) + i \l E_a(x).  & (2.3)}
$$

The diffeomorphism $\f_\l$ allows us to take across group actions
from the right- to the left-hand side of (2.3) and vice versa.
The $G$-gauge transformations (2.2) acting on the right-hand side
of (2.3) have the form

$$
A^\C_a(x) \mapsto
  g^{-1}(x) A^\C_a(x) g(x) + g^{-1}(x) \partial_a g(x) \eqno(2.4)
$$

\ni with $g(x) \in G$.
Consider the Lie algebra $Lie(G^\C) ={Lie (G)}^\C = {Lie (G)} + i\,
Lie (G)$ of $G^\C$, where

$$
G^\C  = G\ e^{i\, {Lie (G)} }      \eqno(2.5)
$$

\ni is the (unique for a compact connected Lie group $G$)
universal complexification of the group $G$ (see, for example,
[8]).
In particular for $G = SU(N)$, its complexification $SU(N)^\C$
is (group theoretically) isomorphic to $SL(N, \C), \ N \geq 2$.
Since $A_a^\C(x)$ is a one-form with values in ${Lie (G)}^\C$, we
may regard it as a $G^\C$-connection and define on it
$G^\C$-transformations given by the following
obvious extension of (2.4),

$$
A^\C_a(x) \mapsto  A^{\C}{}'_a(x) =
  g_\C^{-1}(x) A^\C_a(x) g_\C(x) + g_\C^{-1}(x) \partial_a g_\C(x),
\eqno(2.6)
$$

\ni where $g_\C(x) \in G^\C$. The $\l$-dependent action induced
on the left-hand side of (2.3) reads

$$
\eqalign{
A_a(x) &\mapsto  Re( A^{\C}{}'_a(x))  \cr
\tilde E^a(x) &\mapsto  { \sqrt{h} h^{ab}(x)  \over \l}
\, Im(A^{\C}{}'_b(x)) } \eqno(2.7)
$$

\ni or, under an infinitesimal change generated by the algebra element
$\xi(x) +i \eta(x) \in {Lie (G)}^\C$,

$$
\eqalign{
\d A_a(x) & =   D_a \xi(x) - \l [E_a(x), \eta(x)]  \cr
\l \d \tilde E^a(x) & = \l [\tilde E^a(x), \xi(x)] +
\tilde D^a \eta (x).} \eqno(2.8)
$$

Hence the identification of the big configuration space
of the complex $G^\C$-Yang-Mills theory with the big phase space
of the real $G$-Yang-Mills theory via (2.3) allows us to introduce in the
latter an action of the group  of local $G^\C$-transformations
(one action for each choice of $\l, \ \l \in \R$).
Note that the action (2.7) on $\cal P$ is not symplectic, only the subgroup
of ``real" gauge transformations is.

By introducing a larger group ${\cal G}^\C$ as a group of gauge transformation
on these spaces one may a priori worry that requiring
invariance with respect to the ${\cal G}^\C$-action leads to a loss of
relevant physical observables. The result we obtain in the following
together with the result of [5] will show that, at least for
 $SU(2)$-Yang-Mills theory, this is not the case.

Recall that physical observables in gauge theory are defined as the
$\cal G$-invariant functions {\it on the constraint
surface} ${\cal C}\subset\cal P$, and not on all of $\cal P$.
We prove that for
every complex ${\cal G}^\C$-orbit ${\cal O}^\C\subset\cal P$ that intersects
the constraint surface $\cal C$, we have

$$
{\cal O}^\C \cap {\cal C} = {\cal O},\eqno(2.9)
$$

\ni where ${\cal O}$ is a single real ${\cal G}$-orbit,
and therefore the intersection contains only one real orbit (corresponding to
a unique physical configuration).
Hence the extra invariance conditions imposed by the
$G^\C$-gauge transformations correspond merely to identifying
unphysical configurations outside the constraint surface $\cal C$, along orbits
transverse to $\cal C$.
This result, together with the known fact
that the $SU(2)^\C$-Wilson loop variables (1.3)
separate (closed) $SL(2, \C)$-gauge orbits in ${\cal P}$ [5], implies
that these variables form a good set of generalized coordinates
on the physical phase space of $SU(2)$-Yang-Mills theory,
i.e. they separate all points in ${\cal P}_{phys}$ (with the
possible exception of singular points for which
the holonomy group is a subgroup of the group of
null rotations [5]).

We now prove our result by contradiction. Consider the
Gauss constraint surface $\cal C$ in the big phase space
${\cal P} = T^*{\cal A}$ of the $G$-Yang-Mills theory,

$$
{\cal C} = \lbrace (A_a, \tilde E^a) \in {\cal P} \mid
D_a\tilde E^a = 0 \rbrace.     \eqno(2.10)
$$

\ni Assume that two different ${\cal G}$-orbits ${\cal O}^{(0)}$, ${\cal
O}^{(1)}$ in ${\cal C}$ are contained in the same ${\cal G}^\C$-orbit
and let

$$
p_j = (A^{(j)}_a, \tilde E^{(j)a});  \quad  j = 0, 1    \eqno(2.11)
$$

\ni denote two points in these orbits. Then by assumption
there exists a complex gauge transformation
connecting them, which according to (2.5) is of the form

$$
g_0(x)\, e^{i \xi_0(x)} ,
$$

\ni where $g_0(x) \in {\cal G}, \xi_0(x) \in {Lie ({\cal G})}$. Without
loss of generality we can assume that $g_0(x) = e$.
Indeed let $g_0(x)$ be nontrivial. Then the point $p'_1 = e^{i \xi_0(x)}
p_0$ is in the same ${\cal G}$ orbit as $p_1$ ($p_1 = g_0(x)
e^{i \xi_0(x)}
p_0  = g_0(x) p'_1$) and we can prove our result for the pair $(p_0, p'_1)$.

Consider the curve

$$
q{(s)} = (A^{(s)}_a, \tilde E^{(s)a})  = e^{i s \xi_0} p_0
$$

\ni in phase space, with $q(0) = p_0$ and $q(1) = p_1$. We will show that
necessarily $p_0 = p_1$, which contradicts
the initial assumption that these points belong to different
real orbits.

Consider the following real function on the unit interval:
\note{Our proof is the extension
to infinite-dimensional Yang-Mills systems of analogous proofs
used in finite-dimensional systems. The geometric interpretation
of our construction and finite-dimensional examples are discussed
elsewhere [9].}

$$
 r(s) = \int_\Sigma dx\, Tr \left\{ \left( \partial_a \tilde E^{(s)a}(x) +
\left[ A^{(s)}_a(x), \tilde  E^{(s)a}(x) \right] \right) \xi_0(x)
\right\}. \eqno(2.12)
$$

\ni for which we have

$$
r(0) = r(1) = 0.   \eqno(2.13)
$$

For the derivative of $r(s)$ with respect to the curve parameter $s$ we obtain
(using (2.8)) the following non-negative expression

$$
\eqalign{
\dot r(s) = - \int_\Sigma \sqrt{ h}\, Tr & \left( \frac{ h^{ab}}{\l}
    \left(D[A^{(s)}]_a \xi_0 \right)(x)\left(D[A^{(s)}]_b \xi_0\right)(x)
\right. \cr
 + & \left. \frac{\l}{\sqrt{h}}\, h_{ab}
\left[ \tilde E^{(s)a}(x), \xi_0(x) \right]
\left[ \tilde E^{(s)b}(x), \xi_0(x) \right] \right) \geq 0,}
\eqno(2.14)
$$

\ni where we have assumed appropriate boundary conditions on the fields to
ensure the vanishing of boundary terms. Note
that we are using a normalization for the
$G$-generators $\tau^i$ for which
$Tr\,\tau^i \tau^j =-\frac{1}{2}\delta^{ij}$. Combining (2.13) with (2.14) we
conclude that

$$
r(s) \equiv \dot r(s) \equiv 0, \quad s \in [0, 1].
$$

\ni (Note that the same argument would not be valid for non-compact gauge
groups $G$, since in that case the inner product on the Lie algebra used in
(2.14) would have a different signature.) This implies that the vector

$$
\left( \d_{\xi_o} A^{(s)}_a,  \d_{\xi_o} \tilde E^{(s)a} \right) =
\left( (D[A^{(s)}]_a \xi_0),
[ \tilde E^{(s)a}, \xi_0 ] \right)\eqno(2.15)
$$

\ni and therefore also the vector

$$
\left( \d_{i \xi_o} A^{(s)}_a,  \d_{i \xi_o} \tilde E^{(s)a} \right)=
\left(- \l
[  E^{(s)}_a, \xi_0 ],
 {1 \over \l} \widetilde {D[A^{(s)}]}^a \xi_0 \right)\eqno(2.16)
$$

\ni vanish at any point $q(s) = ( A^{(s)}_a,  \tilde E^{(s)a})$,
$s \in [0, 1]$. Since (2.16) are the infinitesimal ``imaginary"
transformations generated by $i \xi_0$ we conclude that

$$
( A^{(0)}_a,  \tilde E^{(0)a}) =
( A^{(1)}_a,  \tilde E^{(1)a}).   \eqno(2.17)
$$

We have therefore proven our claim that no
additional conditions are imposed by requiring invariance under the
${\cal G}^\C$-action on $\cal P$. The following comment is in order:

\item{} Equations (2.12) and (2.14) show that locally the imaginary gauge
transformations are transverse to the constraint surface. This
is in accordance with the Moncrief decomposition of the tangent
space at every point of the constraint surface on the phase space of
Yang-Mills systems [10].

\vskip 1.5cm

\line{\ch 3. The loop algebra\hfil}

It follows from the previous section that the complex $SU(2,\C)$-Wilson loop
variables (1.3) may serve as an alternative to the usually employed set of
generalized coordinates

$$
T^{(n)}_\a{}_{x_1 x_2\dots x_n}^{a_1 a_2\dots a_n}(A,\tilde E)=Tr\,
\tilde E^{a_1}(x_1)U_\a(x_1,x_2)\tilde E^{a_2}(x_2)U_\a(x_2,x_3)\dots
\tilde E^{a_n}(x_n)U_\a(x_n,x_1) \eqno(3.1)
$$

\ni on the $SU(2)$-Yang-Mills phase space [3, 11]. In (3.1), $U_\a(x_i,x_j)
 = U_\a(A)(x_i,x_j) $ denotes
the holonomy along $\a$, taken between the two points $x_i$ and
$x_j$, $U_\a(x_i,x_j)={\rm P}\exp\int_{x_i}^{x_j}A_a\, dx^a$.
We may expand the complex Wilson loop (1.3) as a power series in $\l$,

$$
T_\a(A+i\l E)=\frac12\,\sum_{n=0}^{\infty}\frac{(i\l )^n}{n!}\, T^n(\a,A,E),
\eqno(3.2)
$$

\ni where

$$
\eqalign{
T^n(\a,A,E)=Tr\, & \int_0^1 dt_1 \int_{t_1}^1 dt_2\dots \int_{t_{n-1}}^1 dt_n\,
U(\a(0),\a(t_1))E_{a_1}(\a (t_1))
  \dot\a^{a_1}(t_1)   U(\a(t_1),\a(t_2)) \cr
& E_{a_2}(\a (t_2)) \dot\a^{a_2}(t_2)
\dots E_{a_n}(\a (t_n))\dot\a^{a_n}(t_n) U(\a(t_n),\a(1)), \cr}\eqno(3.3)
$$

\ni are nothing but integrated versions of the $T^{(n)}$ in (3.1).
Hence the $T_\a(A+i\l E)$ may be regarded as generating functions for the
``higher Wilson loop momenta" $T^n$. (A similar expansion is used in the
zero-curvature formulation
of reduced gravitational models [12].) We may now
explicitly calculate the Poisson bracket between two complex Wilson
 loops, using the fundamental relation (2.1):

$$
\eqalign{
\{ T_\a&(A+i\l E), T_{\a'}(A+i\l' E) \}=\cr
&i (\l'-\l )\,  S[\a,\a']  \sum_{l=1}^3 Tr \Big[ U_\a(A+i\l E)(\a(t),\a(t))
\tau^l
\Big] Tr \Big[ U_{\a'}(A+i\l' E)(\a'(t'),\a'(t')) \tau^l \Big] \ .
}\eqno(3.4)
$$

\ni The path-dependent, distributional (for $d>2$) structure constants are
defined by

$$
S[\a,\a']=\frac{1}{\sqrt{h}}\int_0^1 dt \int_0^1 dt'\; \delta^d(\a
(t),\a'(t')) h_{ab}(\a(t))\dot\a^a(t)\dot\a'{}^b(t'),\eqno(3.5)
$$

\ni and vanish for perpendicular tangent vectors $\dot\a$, $\dot\a'$. The
right-hand side of (3.4) may be evaluated further by using the identities for
the
$\tau$-matrix generators of the $su(2)$-algebra.

The cases of special interest are those where $\l' =\l$, for which the Poisson
brackets vanish, and $\l'= -\l$, for which the connection arguments are
canonically conjugate. For the latter case, the right-hand side of (3.4)
may again be expanded as a power series in $\l$. However, it is not true that
the right-hand side is expressible as (a sum of terms) $T_\g(A\pm i\l E)$ for
some loop(s) $\g$. Rather, one obtains Wilson loop variables with a
``colouring" for pieces of loops between intersections, which keeps track of
whether the holonomy for that piece comes from an integration of the connection
$A+i\l E$ or of its conjugate $A-i\l E$.

The idea of using the complex Wilson loops $T_\a(A+ i \l E)$ as generating
functions for Wilson loop momenta may now profitably be used to define natural
momenta $T^n(\a,A,E)$ for the case when $\a$ possesses self-intersections.
For simple (i.e. non-selfintersecting) loops we keep the definition (3.3).
Then,
by taking Poisson brackets as in (3.4), with both $\a$ and $\a'$ simple,
we may define $T^n(\a,\a',A,E)$ as the coefficient at order
$\l^{n+1}$, which is a finite sum of terms. For loop configurations with more
complicated intersections, we can continue to take Poisson brackets of the
resulting quantities.
This is a bona fide procedure from the point of view of the
real gauge theory, since the Poisson bracket of two $SL(2,\C)$-invariant
quantities is always invariant under $SU(2)$- (though generally not under
$SL(2,\C)$-)gauge transformations.
The reason for this is that the $SU(2)$-gauge transformations are
canonical (i.e. Poisson bracket preserving)
transformations while the $SL(2, \C)$ gauge transformations are not.

\vskip1.5cm

\line{\ch 4. Conclusions\hfil}

We have shown that the complex Wilson loops $T_\a(A\pm i \l E)$ form a
good set
of generalized coordinates on the phase space of the real $SU(2)$-Yang-Mills
theory. If the result of Ashtekar and Lewandowski [5] extends to any
compact gauge
group $G$, also our result immediately generalizes. This shows that
there is a sense in which the identification $(A,E)\leftrightarrow A+i \l E$
between
$T^*\cal A$ and ${\cal A}^\C$ continues to hold at the level of the
corresponding physical spaces $T^*({\cal A}/{\cal G})$ and ${\cal A}^\C
/{\cal G}^\C$.

This is a non-trivial result because of the non-linear character of the
quotient spaces involved. Note that we have not shown that each point
in ${\cal A}^\C/{\cal G}^\C$ does in fact correspond to a physical phase
space point. There are finite-dimensional gauge model systems for which one can
prove a one-to-one correspondence between a dense subset of  ${\cal A}^\C
/{\cal G}^\C$ and $T^*({\cal A}/{\cal G})$ [9], but there is as yet no
proof for the infinite-dimensional (non-abelian) gauge theory case.

It remains to be investigated how our alternative Hamiltonian description for
Yang-Mills theory intertwines with the dynamical evolution, and whether it
leads to any simplification in an explicitly gauge-invariant description, for
example, in a regularized lattice formulation. More generally, since our result
is of a kinematical nature, it may be applied to any theory whose
configuration space is a space of connections, for example, a Chern-Simons
theory [13].

As already mentioned in the introduction, the big phase space of general
relativity in the Ashtekar formulation is ${\cal A}^{SL(2, \C)}$, but with a
symplectic structure significantly more complicated than the one of
$SU(2)$-Yang-Mills theory (see [2]).
Notice that the fact that $A^\C_{grav}$ is a (holomorphic)
coordinate on the phase space of general relativity is clear from (1.1)
or from the reality conditions
$\overline A_{grav}^\C = - A_{grav}^\C + 2 \Gamma (E)$.
We hope that techniques similar to the
ones employed in the present paper will be useful in establishing an
analogous rigorous result in general relativity.

\vskip0.8cm

\vbox{

\ni {\bf Acknowledgements}

It is a pleasure to thank A. Ashtekar, J. Lewandowski, R. Picken
and T. Thiemann for
useful discussions.
This work was supported partially by funds provided by the PSU.
JMM was also supported by NATO grant 9/C/93/PO and by  C.E.C.
project SCI-CT91-0729.
      }

\vskip1cm

\line{\ch References\hfil}

\item{[1]} A. Ashtekar, {\it Phys. Rev. Lett.} {\bf 57} (1986) 2244;
{\it Phys. Rev. D} {\bf 36} (1987) 1587

\item{[2]}A. Ashtekar, {\it Non-perturbative canonical quantum gravity}
(notes prepared in collaboration with R.S. Tate), World Scientific,
Singapore, 1991

\item{[3]} C. Rovelli and L. Smolin, {\it Nucl. Phys.} {\bf B331} (1990)
80

\item{[4]} B. Br\"{u}gmann, R. Gambini and J. Pullin,  {\it
Phys. Rev. Lett.} {\bf 68} (1992) 431

\item{[5]} A. Ashtekar and J. Lewandowski, {\it Class. Quant. Grav.}
{\bf 10} (1993) L69

\item{[6]} S. Mandelstam, {\it Phys. Rev.} {\bf 175} (1968) 1580

\item{[7]} R. Loll, {\it Nucl. Phys.} {\bf B368} (1992) 121; {\bf B400}
(1993) 126

\item{[8]} G. Hochschild,  {\it The structure of Lie groups},
Halden-Day, San Francisco, 1965

\item{[9]} R. Loll, J.M. Mour\~ao and J.N. Tavares,  {\it
Complexification of gauge theories}, to appear in {\it J. Geom. Phys.}

\item{[10]} V. Moncrief, {\it J. Math. Phys.} {\bf 16}(1975) 1556

\item{[11]} R. Gambini and A. Trias, {\it Nucl. Phys.} {\bf B278} (1986) 436

\item{[12]} V. Husain, {\it Phys. Rev. D} {\bf 55} (1994) 6207

\item{[13]} R. Loll, J.M. Mour\~ao and J.N. Tavares, in {\it
Constraint theory and quantization methods}, ed. F. Colomo, L. Lusanna
and G. Marmo, World Scientific, Singapore, 1994, 291

\vfill

\eject

\bye